\documentclass[3p,times,twocolumn]{article}
\usepackage{amssymb}
\usepackage[figuresright]{rotating}
\begin{document}

\title{Absolute Frequency Atlas from 915~nm to 985~nm based on Laser Absorption Spectroscopy of Iodine}

\author{Christian N\"olleke\thanks{Corresponding author: christian.noelleke@toptica.com}, Christoph Raab, Rudolf Neuhaus, Stephan Falke \\ \\TOPTICA Photonics AG,  Lochhamer Schlag 19, 82166 Gr\"afelfing, Germany}

\date{}

\maketitle

\begin{abstract}
This article reports on laser absorption spectroscopy of iodine gas between 915~nm and 985~nm. This wavelength range is scanned utilizing a narrow linewidth and mode-hop-free tunable diode-laser whose frequency is actively controlled using a calibrated wavelength meter. This allows us to provide an iodine atlas that contains almost 10,000 experimentally observed reference lines with an uncertainty of 50 MHz. For common lines, good agreement is found with a publication by Gerstenkorn et al. \cite{gerstenkorn1978}. The new rich dataset allows existing models of the iodine molecule to be refined and can serve as a reference for laser frequency calibration and stabilization.\end{abstract}

\section{Introduction}
Frequency standards are of prime importance in modern metrology applications. The most precise standards are realized by optical clocks, which --- through laser spectroscopy and frequency stabilization of the interrogating laser field to a narrow clock transition ---  already create frequency standards beyond the best realizations of the SI definition of the second. Thus frequency standards with relative frequency instabilities below the representations of the SI second derived in microwave clocks are created \cite{ludlow2015}. Frequency combs are used to link between the microwave and the optical frequency domain: The precision and the stability of one frequency standard may be transferred to another. For example, the frequency of a cw laser may be either counted or stabilized relative to the timing signals of GPS \cite{kliese2016, friebe2008, peters2015}. However, frequency combs are too elaborate for many experiments. A simpler and low-cost alternative is the use of a laser that is stabilized to known transition frequencies of atoms or molecules. Such a setup provides stabilities of a fraction of the natural linewidth, which is typically a few MHz. For example, in cold-atom experiments the lasers for cooling the atoms are frequency stabilized to a saturated absorption spectroscopy signal using an appropriate reference gas cell \cite{preston1996}. In many cases a direct reference is not available: Lasers driving repumping transitions cannot be stabilized because none of the connected states are thermally populated. Another example is a multi-photon excitation scheme, e.g. to excite Rydberg states. Further examples include lasers cooling trapped ions or lasers that need to be stabilized to a frequency defined by an effect on a transition such as a magic wavelength in an optical lattice clock \cite{katori2003, ye2008} or in spinor gases \cite{lundblad2010}. To overcome this problem, wavelength meters can be used. Commercial instruments, e.g. based on Fizeau interferometers, can provide precise frequency readings down to the MHz level, however require a reference laser of known frequency for calibration to reach this accuracy.

Rotational and vibrational levels in molecules provide a rich manifold of absorption lines over huge spectral ranges. In particular iodine can serve as a frequency standard throughout the visible and near infrared region \cite{Bodermann2000}. Hyperfine transitions of iodine within the gain profile of helium-neon lasers serve as secondary representation of the SI-second \cite{bipm2016}. Other frequencies of specific interest have triggered highly accurate frequency measurements of particular iodine lines \cite{kobayashi2016, ye1999}. Another application example of iodine cells serving as a frequency reference is the search for exoplanets: Iodine absoption spectra are co-recorded with the spectra of stars whose lines are frequency shifted by the Doppler effect due to the movement induced by an orbiting planet \cite{schwab2012, yilmaz2017}.

Iodine also provides a wealth of spectroscopic data that has been used to derive models of the molecular states whose ro-vibrational (and hyperfine) levels provide seemingly neverending lists of transition frequencies. The  B$\,^3\Pi^+_{0u}$ -- X$\,^1\Sigma_g^+$ system of iodine has been analyzed and modeled in great detail and with high precision \cite{salami2005}, including hyperfine structure for the different isotope combinations \cite{knockel2004, salumbides2008}. Precision spectroscopy of the B -- X transitions enables laser frequency stabilization on the level of MHz or even below \cite{simonsen1997, edwards1999}. The A$\,^3\Pi_{1u}$ -- X$\,^1\Sigma_g^+$ system has a much smaller coupling strength resulting in high power requirement for saturation of the transitions. Hence only Doppler broadened spectroscopy data is available \cite{gerstenkorn1978}. The latest analysis is based on data from Fourier transform spectroscopy of light scattered from iodine molecules \cite{appadoo1996}. In this work, we take another experimental approach: The light of a widely-tunable laser is sent through a glass cell filled with iodine gas. We compose a list of iodine absorption lines, which can be used as a frequency standard in the spectral range between 915~nm and 985~nm. This atlas of transition frequencies as well as the recorded spectrum are available through the online data repository of this article \cite{atlas_online}.

\section{Experiment}

\subsection{Experimental Setup}
\begin{figure*}[ht]
	\centering
	\includegraphics[width=0.7\textwidth]{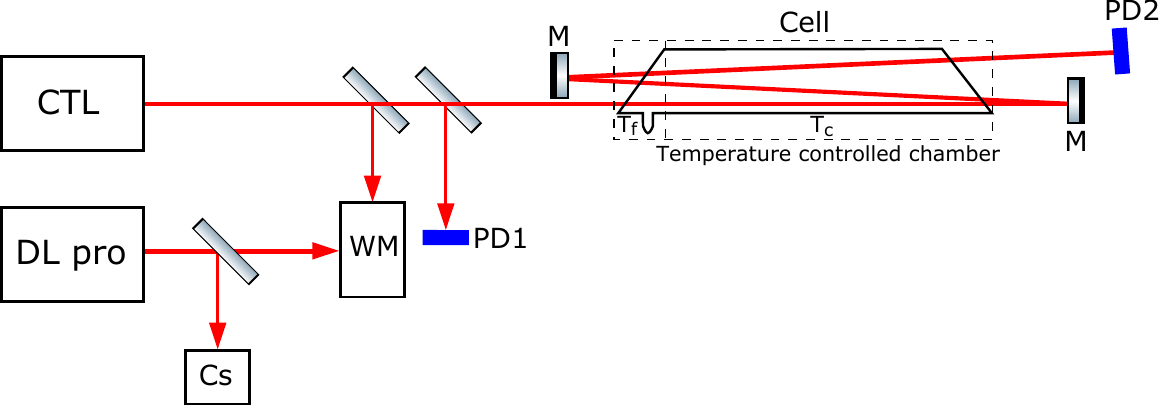}
	\caption{\label{fig:setup}Sketch of the experimental setup. Parts of the beam from the the continuously tunable laser (CTL) are directed to a wavelength meter (WM) and a photodiode (PD1). The beam path is folded using two mirrors (M) such that it passes the cell three times. The cell is located in a temperature-controlled chamber, that allows the temperature of the cell body and the cold finger to be set independently. The transmitted power is measured by a photodiode (PD2). The wavelength meter is calibrated with an external-cavity diode laser (DL~pro) that is stabilized to a Doppler-free cesium spectroscopy (Cs).}
\end{figure*}

Figure \ref{fig:setup} shows the principle of the experimental setup. The linearly polarized light from the laser (TOPTICA DLC CTL 950) is directed to the spectroscopy setup via a single-mode polarization-maintaining fiber thereby also cleaning up the spatial mode. The laser can be tuned from 915~nm to 985~nm without mode hops. The short-term linewidth of the laser is measured using the self-heterodyning method \cite{ludvigsen1994}, and is found to be $<5~\textrm{kHz}$ (corresponding to $<1.5 \times 10^{-5}$~pm) for an integration time of 5~$\mu$s, which is much smaller than the observed spectral features.

A small fraction of the light is tapped-off and directed to a wavelength meter (HighFinesse WS8-2). Another part is directed to an Si-photodiode to measure the power before the cell as reference for determining the relative absorption. The main part of the light passes a 75-cm-long iodine-filled (isotope $^{127}$I, no buffer gas) cell three times, resulting in an effective length of 225~cm. The cell has Brewster-angled windows to minimize reflection losses. The optical intensity inside the cell is approximately $10~\textrm{kW}/\textrm{m}^2$. The power transmitted through the cell is measured with another photodiode. The temperature of the cell body and the cold finger are controlled independently ($T_\mathrm{c}$ and $T_\mathrm{f}$, respectively). We set the cold finger to a temperature of $T_\mathrm{f}=39^\circ \mathrm{C}$, corresponding to an iodine vapour pressure of 127~Pa. The corresponding pressure shift is approximately 1~MHz (0.003~pm) \cite{bodermann1998}. The cell body is heated to $T_\mathrm{c}=300^\circ \mathrm{C}$ in order to increase the signal-to-noise ratio, utilizing the effect of temperature via the Boltzmann distribution on the relative level population.

The laser's vacuum wavelength is continuously measured with the wavelength meter. The scan is performed step-wise. The desired wavelength is set in the wavelength meter's software. A feedback loop to the laser based on a digital PID control then stabilizes the laser wavelength to the value measured by the wavelength meter. The correction rate of the feedback loop is on the order of 10~Hz. After the wavelength has been stabilized, the measurement of the cell transmission is performed. To obtain a high absolute accuracy, the wavelength meter is calibrated to a reference laser with known wavelength. The reference laser is an external-cavity diode laser (TOPTICA DL~pro) stabilized to the F=4 $\rightarrow$ F=4 hyperfine transition of the cesium $D_1$ line (895~nm), which is close to the wavelength range investigated in this study. During the measurements, the wavelength meter is calibrated every five minutes. The drift of the wavelength meter during this interval is negligible. The absolute accuracy of our setup is determined by the accuracy and stability of the reference laser, the performance of the feedback-loop, and the accuracy of the wavelength meter. We find that the major contribution stems from the accuracy of the wavelength meter and its calibration and estimate the uncertainty of the frequency measurement to be 20~MHz (0.06~pm).

\subsection{Spectroscopy results}
\begin{figure}[ht]
	\centering
	\includegraphics[width=\columnwidth]{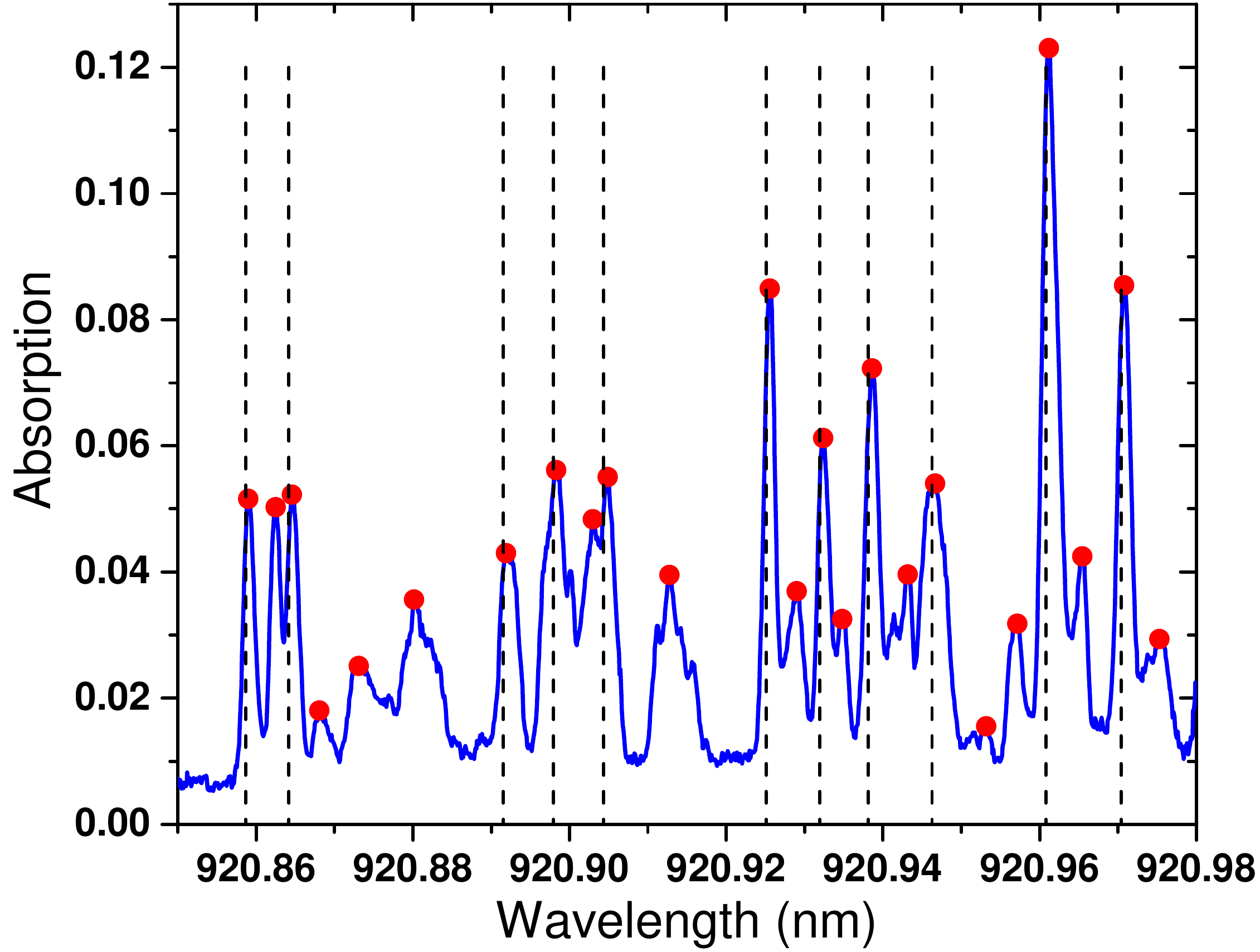}
	\caption{\label{fig:scan1}Small part of the recorded spectrum and determined lines therein. The solid blue line shows the measured spectrum, the red dots indicate identified peaks. The dashed black vertical lines correspond to the lines identified in \cite{gerstenkorn1978}.}
\end{figure}

We perform laser absorption spectroscopy of iodine over the entire spectral range covered by the laser, i.e. from 915~nm to 985~nm, with a scan speed of 1~nm per hour. The wavelength is increased in steps of 0.1~pm (33~MHz) by changing the set wavelength of the feedback loop. For each wavelength, the absorption of the laser by the iodine is measured. The integration time for each data point is 10~ms. As an example, Fig. \ref{fig:scan1} shows a part of the complete data set. The spectral resolution of 0.1~pm is well below the width of the Doppler-broadened lines of 1.5~pm. We note that the chosen step-size is a trade-off between high resolution and measurement time. With the equipment used in this study, the resolution could be as low as 0.003~pm, only limited by the resolution of the wavelength meter. We use the feedback to a piezo electric element acting on the laser cavity whose control has a resolution at the few kHz level ($10^{-5}$~pm). Even smaller frequency steps are possible by controlling the current of the laser diode. The absorption of the strongest lines we observe is 15\%.

We determine the spectral positions and amplitudes of each absorption line by analyzing the recorded spectral data numerically. The employed Python library for peak detection is available online\footnote{https://github.com/kricki/py-findpeaks}. Before conducting the peak analysis, a moving average filter is applied, averaging over five neighboring data points. We use a minimum prominence of 0.3\% as an additional criterium for the peak detection. The spectral position and amplitude are evaluated at the maximum of the line. Due to free space optical paths in the laser absorption spectroscopy setup, absorption lines from water resonances appear in our data induced by humidity of the air. We confirmed in an independent measurement without the iodine cell in the beam path that water lines are present. Since the water spectrum in the near-infrared region is well known, we used an existing database \cite{hitran} to mark all measured lines within 4~pm of a water line (corresponding to the Doppler width of water at room temperature). We only considered those water lines intense enough to be above the signal-to-noise ratio of our measurement by deriving a minimal intensity from comparing the recorded spectrum (without the iodine cell) with the simulated water spectrum. This enables us to ensure that only iodine lines are included in further analysis. We identified a total of 9970 iodine lines in the investigated spectral range. The uncertainty in the absolute frequency of the determined line center measurements of this study is approximately 50~MHz (0.15~pm) and is a combination of the uncertainty in the absolute frequency measurement (20~MHz), the step size of the wavelength scan (33~MHz) and the line position determination (30~MHz).

\subsection{Discussion}
\begin{figure}[ht]
	\centering
	\includegraphics[width=\columnwidth]{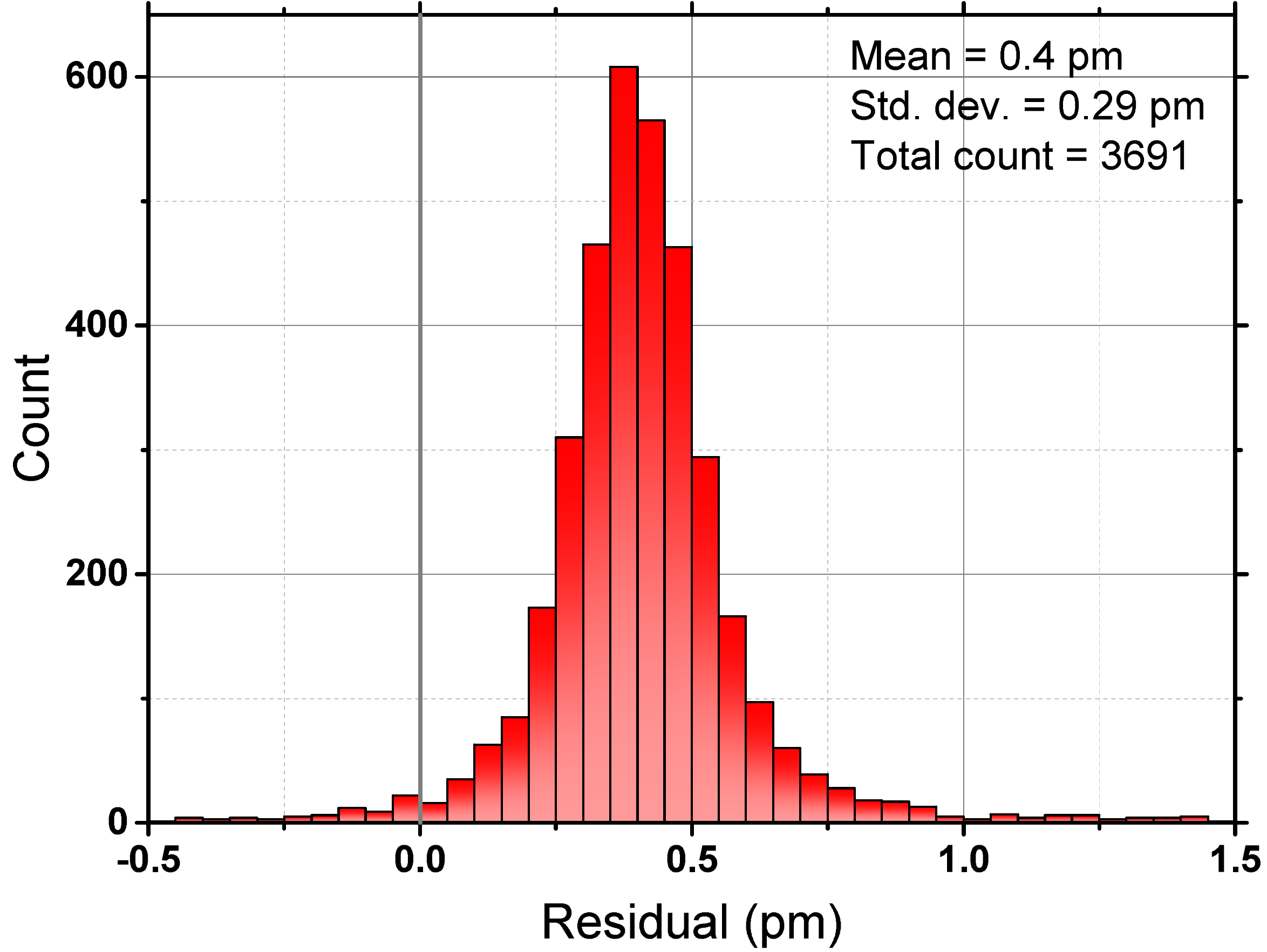}	
	\caption{\label{fig:residuals}Statistics of residuals between this study (accuracy 0.15~pm) and \cite{gerstenkorn1978} (accuracy 0.45~pm).}
\end{figure}

In order to compare the data to the iodine atlas in \cite{gerstenkorn1978}, we calculate the deviation between lines that were identified in \cite{gerstenkorn1978} and in this study. In the investigated spectral range, a total of 3769 lines have been identified in \cite{gerstenkorn1978}. For all but 78 of these lines, we find a corresponding match in our list of identified lines. Missing lines are typically two closely spaced lines that were not identified by the employed algorithm. The statistics of these 3691 residuals is shown in Fig. \ref{fig:residuals}. The mean value of $0.4~\mathrm{pm}$ (corresponding to approx. $130~\mathrm{MHz}$) indicates a systematic offset between the two studies, where the positive sign means a higher measured wavelength in our study. The offset is in agreement with the combined accuracy of these studies, with contributions of 0.45~pm from \cite{gerstenkorn1978} and 0.15~pm from this work. The standard deviation is smaller than the offset, indicating that statistical processes are under good control in both studies.

\begin{figure}[ht]
	\centering
	\includegraphics[width=\columnwidth]{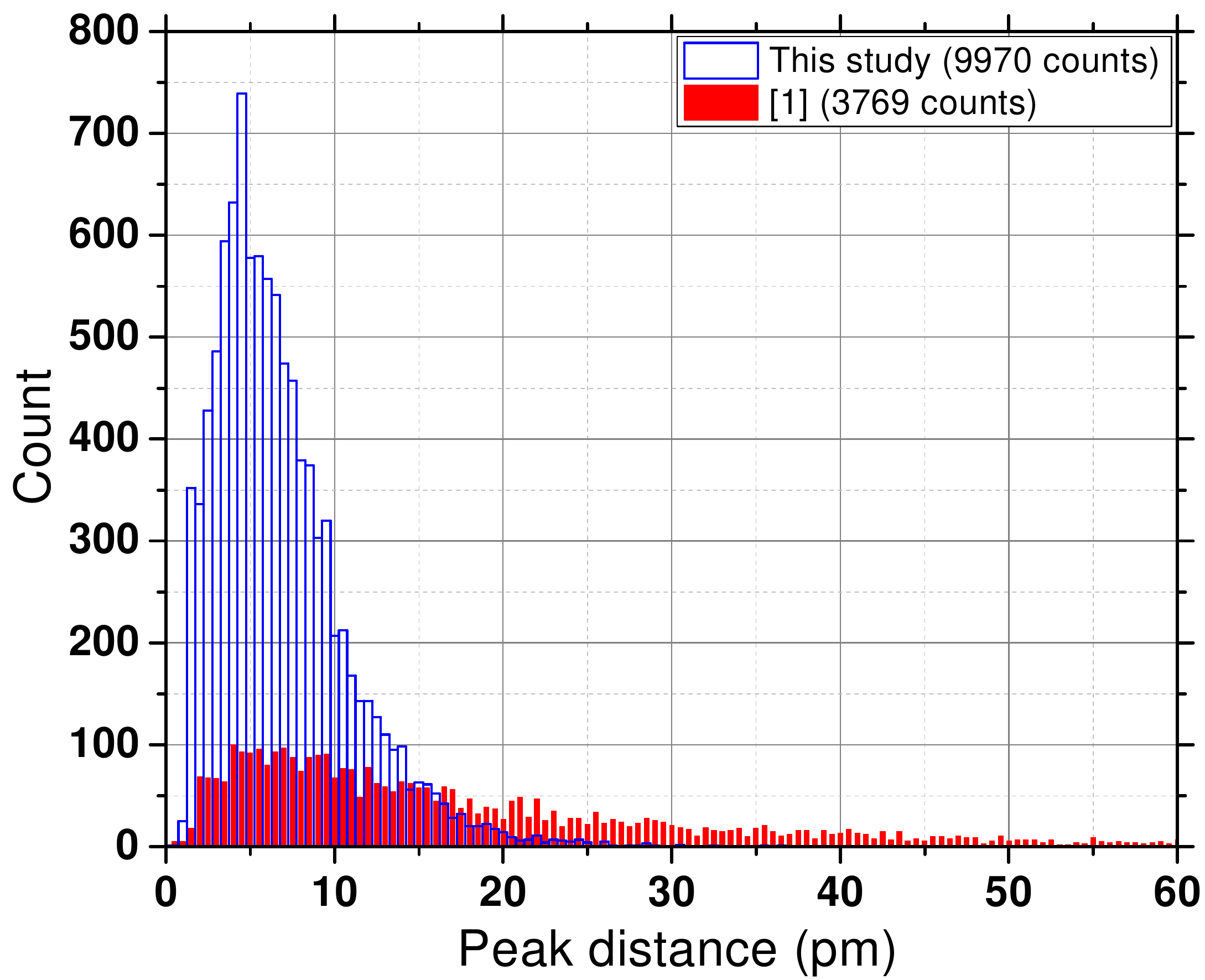}
	\caption{\label{fig:distance_statistics}Statistics of spacing between neighboring lines. Open blue bars show the statistics of this study, red filled bars show the statistics from \cite{gerstenkorn1978} for comparison.}
\end{figure}

Due to the higher spectral resolution of our experimental method, we are able to detect lines that have not been resolved in \cite{gerstenkorn1978}. We identified about 6000 additional lines in the investigated spectral range. As can be seen Fig.~\ref{fig:distance_statistics}, the average spacing between identified spectral lines is approximately 4~\textrm{pm}, compared to about 10~\textrm{pm} in \cite{gerstenkorn1978}. Almost no gap larger than 20~pm between two neighboring line remain in the atlas of lines provided with this study. In contrast, a long tail of large spectral gaps exists in the atlas of \cite{gerstenkorn1978}.

We note that the width of the measured lines is limited by Doppler broadening and the underlying hyperfine structure is not resolved. Hence, a peak of a line does not necessarily align with the transition energy between the ro-vibrational levels. Once lines are assigned to the ro-vibrational levels and the underlying hyperfine structure is understood, fits to the line profile taking the hyperfine components into account can be applied to the primary data \cite{atlas_online}.

\section{Conclusion and Outlook}
This study presents laser absorption spectroscopy of iodine between 915~nm and 985~nm. In this wavelength range, we found almost 10000 absorption lines. We find excellent agreement between our data and the results of a previous study by Gerstenkorn et al. \cite{gerstenkorn1978}, which used an alternative experimental approach. Due to the increased resolution and the improved signal to noise ratio, we were able to add 6000 lines to the existing data file and reduce the frequency uncertainty to 50 MHz. The resulting atlas may serve as reference for spectroscopic experiments in this spectral range with unprecedented line density and frequency accuracy. The improved and enriched data field of this study provided in the data repository of this paper \cite{atlas_online} may help to resolve residual ambiguities of the latest study of the A -- X system.

The resolution and the absolute accuracy of our setup are limited by the accuracy of the wavelength meter. Both can be improved by measuring the laser wavelength with a frequency comb instead of a wavelength meter \cite{fan2014, neuhaus2016} or by dual comb spectroscopy \cite{ideguchi2012}. Conversely, from the known spectral positions of the absorption lines, the CTL laser system in combination with an iodine cell can serve as a synthesizer for optical frequencies with a precision of a few MHz, corresponding to a relative precision of about 1 in $10^8$. The average line spacing of 1$~\mathrm{GHz}$ could be bridged by acousto-optical or electro-optical modulators, so that the laser can be set and stabilized to arbitrary frequencies by referencing it to one of the absorption lines reported in this study.

For more refined theoretical molecular models and frequency standards with an absolute accuracy below $10^{-8}$, Doppler-free spectroscopy is desired. The output power of the laser used in this work is too low to achieve significant saturation, and hence the width of the absorption lines is limited by Doppler broadening. However, using a tunable laser in combination with a tapered amplifier to boost the available power could be used to perform saturation spectroscopy. Alternatively, a build-up cavity would increase the optical power inside the cell.

\section{Acknowledgments}
We thank Horst Kn{\"o}ckel (Leibniz University of Hannover) for providing the iodine cell and Stefan Burgardt and Florian Karlewski (HighFinesse) for providing the wavelength meter.

\bibliographystyle{ieeetr}
\bibliography{references}

\begin{thebibliography}{10}

\bibitem{gerstenkorn1978}
S.~Gerstenkorn and P.~Luc, ``Atlas du spectre d'absorption de la molecule
  d'iode 14800-20000 cm$^{-1}$,'' {\em Paris: Editions du Centre National de la
  Recherche Scientifique (CNRS), 1978}, vol.~1, 1978.

\bibitem{ludlow2015}
A.~D. Ludlow, M.~M. Boyd, J.~Ye, E.~Peik, and P.~O. Schmidt, ``Optical atomic
  clocks,'' {\em Reviews of Modern Physics}, vol.~87, no.~2, p.~637, 2015.

\bibitem{kliese2016}
R.~Kliese, N.~Hoghooghi, T.~Puppe, F.~Rohde, A.~Sell, A.~Zach, P.~Leisching,
  W.~Kaenders, N.~C. Keegan, A.~D. Bounds, {\em et~al.}, ``Difference-frequency
  combs in cold atom physics,'' {\em The European Physical Journal Special
  Topics}, vol.~225, no.~15-16, pp.~2775--2784, 2016.

\bibitem{friebe2008}
J.~Friebe, A.~Pape, M.~Riedmann, K.~Moldenhauer, T.~Mehlst{\"a}ubler,
  N.~Rehbein, C.~Lisdat, E.~M. Rasel, W.~Ertmer, H.~Schnatz, {\em et~al.},
  ``Absolute frequency measurement of the magnesium intercombination transition
  $^1{S}_0 \rightarrow ^3{P}_1$,'' {\em Physical Review A}, vol.~78, no.~3,
  p.~033830, 2008.

\bibitem{peters2015}
E.~Peters, S.~Reinhardt, T.~W. H{\"a}nsch, and T.~Udem, ``Absolute frequency
  and isotope shift of the magnesium $(3s^2) ^1{S}_0 \rightarrow (3s3d)
  ^1{D}_2$ two-photon transition by direct frequency-comb spectroscopy,'' {\em
  Physical Review A}, vol.~92, no.~6, p.~063403, 2015.

\bibitem{preston1996}
D.~W. Preston, C.~E. Wieman, and K.~M. Siegbahn, ``Doppler-free saturated
  absorption spectroscopy: Laser spectroscopy,'' {\em Am. J. Phys}, vol.~64,
  no.~11, pp.~1432--1436, 1996.

\bibitem{katori2003}
H.~Katori, M.~Takamoto, V.~Pal{'}chikov, and V.~Ovsiannikov, ``Ultrastable
  optical clock with neutral atoms in an engineered light shift trap,'' {\em
  Physical Review Letters}, vol.~91, no.~17, p.~173005, 2003.

\bibitem{ye2008}
J.~Ye, H.~Kimble, and H.~Katori, ``Quantum state engineering and precision
  metrology using state-insensitive light traps,'' {\em science}, vol.~320,
  no.~5884, pp.~1734--1738, 2008.

\bibitem{lundblad2010}
N.~Lundblad, M.~Schlosser, and J.~Porto, ``Experimental observation of
  magic-wavelength behavior of $^{87}${Rb} atoms in an optical lattice,'' {\em
  Physical Review A}, vol.~81, no.~3, p.~031611, 2010.

\bibitem{Bodermann2000}
B.~Bodermann, M.~Klug, U.~Winkelhoff, H.~Kn{\"o}ckel, and E.~Tiemann, ``Precise
  frequency measurements of lines in the near infrared by {Rb} reference
  lines,'' {\em The European Physical Journal D}, vol.~11, pp.~213--225, jul
  2000.

\bibitem{bipm2016}
BIPM, ``Recommended values of standard frequencies,'' 2016.

\bibitem{kobayashi2016}
T.~Kobayashi, D.~Akamatsu, K.~Hosaka, H.~Inaba, S.~Okubo, T.~Tanabe, M.~Yasuda,
  A.~Onae, and F.-L. Hong, ``Absolute frequency measurements and hyperfine
  structures of the molecular iodine transitions at 578 nm,'' {\em JOSA B},
  vol.~33, no.~4, pp.~725--734, 2016.

\bibitem{ye1999}
J.~Ye, L.~Robertsson, S.~Picard, L.-S. Ma, and J.~L. Hall, ``Absolute frequency
  atlas of molecular {I}$_2$ lines at 532~nm,'' {\em IEEE Transactions on
  Instrumentation and Measurement}, vol.~48, no.~2, pp.~544--549, 1999.

\bibitem{schwab2012}
C.~Schwab, T.~Gutckeb, J.~F. Sproncka, D.~A. Fischera, and A.~Szymkowiaka,
  ``Investigating spectrograph design parameters with the {Yale Doppler}
  diagnostic facility,'' in {\em Proc. of SPIE Vol}, vol.~8446, pp.~844695--1,
  2012.

\bibitem{yilmaz2017}
M.~Yilmaz, B.~Sato, I.~Bikmaev, S.~Selam, H.~Izumiura, V.~Keskin, E.~Kambe,
  S.~Melinkov, A.~Galeev, I.~Ozavci, {\em et~al.}, ``A {Jupiter}-mass planet
  around the {K0} giant {HD 208897},'' {\em Astronomy \& Astrophysics}, 2017.

\bibitem{salami2005}
H.~Salami and A.~J. Ross, ``A molecular iodine atlas in ascii format,'' {\em
  Journal of Molecular Spectroscopy}, vol.~233, no.~1, pp.~157--159, 2005.

\bibitem{knockel2004}
H.~Kn{\"o}ckel, B.~Bodermann, and E.~Tiemann, ``High precision description of
  the rovibronic structure of the {I}$_2$ {B-X} spectrum,'' {\em The European
  Physical Journal D-Atomic, Molecular, Optical and Plasma Physics}, vol.~28,
  no.~2, pp.~199--209, 2004.

\bibitem{salumbides2008}
E.~J. Salumbides, K.~S. Eikema, W.~Ubachs, U.~Hollenstein, H.~Kn{\"o}ckel, and
  E.~Tiemann, ``Improved potentials and {B}orn-{O}ppenheimer corrections by new
  measurements of transitions of $^{129}${I}$_2$ and $^{127}${I} $^{129}${I} in
  the {B}${^3} {\Pi}_{0_u^+} - {X}^1 {\Sigma}_g^+$ band system,'' {\em The
  European Physical Journal D}, vol.~47, no.~2, pp.~171--179, 2008.

\bibitem{simonsen1997}
H.~R. Simonsen, ``Iodine-stabilized extended cavity diode laser at
  $\lambda=633$~nm,'' {\em IEEE transactions on instrumentation and
  measurement}, vol.~46, no.~2, pp.~141--144, 1997.

\bibitem{edwards1999}
C.~Edwards, G.~Barwood, P.~Gill, and W.~Rowley, ``A 633 nm iodine-stabilized
  diode-laser frequency standard,'' {\em Metrologia}, vol.~36, no.~1, p.~41,
  1999.

\bibitem{appadoo1996}
D.~Appadoo, R.~Le~Roy, P.~Bernath, S.~Gerstenkorn, P.~Luc, J.~Verges,
  J.~Sinzelle, J.~Chevillard, and Y.~d’Aignaux, ``Comprehensive analysis of
  the {A--X} spectrum of {I}$_2$: An application of near-dissociation theory,''
  {\em The Journal of chemical physics}, vol.~104, no.~3, pp.~903--913, 1996.

\bibitem{atlas_online}
C.~N{\"o}lleke, C.~Raab, R.~Neuhaus, and S.~Falke, ``Laser absorption
  spectroscopy of iodine between 915 and 985~nm,'' {M}endeley {D}ata, v1,
  http://dx.doi.org/10.17632/cvr6vv2zps.1, 2017.

\bibitem{ludvigsen1994}
H.~Ludvigsen and E.~B{\o}dtker, ``New method for self-homodyne laser linewidth
  measurements with a short delay fiber,'' {\em Optics communications},
  vol.~110, no.~5-6, pp.~595--598, 1994.

\bibitem{bodermann1998}
B.~Bodermann, {\em Untersuchung zur {R}ealisierung eines durchstimmbaren,
  hochpr{\"a}zisen {F}requenzstandards im {NIR} und zur {E}rweiterung des
  {S}pektralbereiches mit {H}ilfe des $^{127}${I}-{M}olek{\"u}ls}.
\newblock PhD thesis, University of Hannover, 1998.

\bibitem{hitran}
L.~S. Rothman, I.~E. Gordon, Y.~Babikov, A.~Barbe, D.~C. Benner, P.~F. Bernath,
  M.~Birk, L.~Bizzocchi, V.~Boudon, L.~R. Brown, {\em et~al.}, ``The
  {HITRAN2012} molecular spectroscopic database,'' {\em Journal of Quantitative
  Spectroscopy and Radiative Transfer}, vol.~130, pp.~4--50, 2013.

\bibitem{fan2014}
I.~Fan, C.-Y. Chang, L.-B. Wang, S.~L. Cornish, J.-T. Shy, and Y.-W. Liu,
  ``Refined determination of the muonium-deuterium {1S-2S} isotope shift
  through improved frequency calibration of iodine lines,'' {\em Physical
  Review A}, vol.~89, no.~3, p.~032513, 2014.

\bibitem{neuhaus2016}
R.~Neuhaus, F.~Rohde, E.~Benkler, T.~Puppe, C.~Raab, R.~Unterreitmayer,
  A.~Zach, H.~R. Telle, and J.~Stuhler, ``1 {THz} synchronous tuning of two
  optical synthesizers,'' in {\em SPIE Photonics Europe}, pp.~99001E--99001E,
  International Society for Optics and Photonics, 2016.

\bibitem{ideguchi2012}
T.~Ideguchi, A.~Poisson, G.~Guelachvili, T.~W. H{\"a}nsch, and N.~Picqu{\'e},
  ``Adaptive dual-comb spectroscopy in the green region,'' {\em Optics
  letters}, vol.~37, no.~23, pp.~4847--4849, 2012.

\end{thebibliography}

\end{document}